\documentclass[12pt]{article}
\usepackage{amsmath}
\usepackage{graphicx}
\usepackage{enumerate}
\begin{document}

\def\a{\alpha}
\def\b{\beta}
\def\c{\varepsilon}
\def\d{\delta}
\def\e{\epsilon}
\def\f{\phi}
\def\g{\gamma}
\def\h{\theta}
\def\k{\kappa}
\def\l{\lambda}
\def\m{\mu}
\def\n{\nu}
\def\p{\psi}
\def\q{\partial}
\def\r{\rho}
\def\s{\sigma}
\def\t{\tau}
\def\u{\upsilon}
\def\v{\varphi}
\def\w{\omega}
\def\x{\xi}
\def\y{\eta}
\def\z{\zeta}
\def\D{\Delta}
\def\G{\Gamma}
\def\H{\Theta}
\def\L{\Lambda}
\def\F{\Phi}
\def\P{\Psi}
\def\S{\Sigma}

\def\o{\over}
\def\beq{\begin{eqnarray}}
\def\eeq{\end{eqnarray}}
\def\lsim{\mathrel{\rlap{\lower 4pt \hbox{$\sim$}}\raise 2pt \hbox{$<$}}}
\def\gsim{\mathrel{\rlap{\lower 4pt \hbox{$\sim$}}\raise 2pt \hbox{$>$}}}
\newcommand{\vev}[1]{ \left\langle {#1} \right\rangle }
\newcommand{\bra}[1]{ \langle {#1} | }
\newcommand{\ket}[1]{ | {#1} \rangle }
\newcommand{\EV}{ {\rm eV} }
\newcommand{\KEV}{ {\rm keV} }
\newcommand{\MEV}{ {\rm MeV} }
\newcommand{\GEV}{ {\rm GeV} }
\newcommand{\TEV}{ {\rm TeV} }
\def\diag{\mathop{\rm diag}\nolimits}
\def\Spin{\mathop{\rm Spin}}
\def\SO{\mathop{\rm SO}}
\def\O{\mathop{\rm O}}
\def\SU{\mathop{\rm SU}}
\def\U{\mathop{\rm U}}
\def\Sp{\mathop{\rm Sp}}
\def\SL{\mathop{\rm SL}}
\def\tr{\mathop{\rm tr}}

\def\IJMP{Int.~J.~Mod.~Phys. }
\def\MPL{Mod.~Phys.~Lett. }
\def\NP{Nucl.~Phys. }
\def\PL{Phys.~Lett. }
\def\PR{Phys.~Rev. }
\def\PRL{Phys.~Rev.~Lett. }
\def\PTP{Prog.~Theor.~Phys. }
\def\ZP{Z.~Phys. }


\baselineskip 0.7cm

\begin{titlepage}

\begin{flushright}
UT-07-33
\end{flushright}

\vskip 1.35cm
\begin{center}
{\large \bf

Inverted Hybrid Inflation as a solution to gravitino problems in Gravity
 Mediation.

}
\vskip 1.2cm
H. Nakajima${}^{1}$ and  Y.  Shinbara${}^{1}$
\vskip 0.4cm

${}^1${\it Department of Physics, University of Tokyo,\\
     Tokyo 113-0033, Japan}

\vskip 1.5cm

\abstract{
  It was recently found that the decay of inflaton and the SUSY breaking
 field produces many gravitinos in the gravity
 mediation scenario. These discoveries led to an exclusion of many
 inflation models such as chaotic, (smooth) hybrid, topological and
 new inflation models.
  Under these circumstances we searched for a successful inflation model and
 found that the ``inverted'' hybrid inflation models
  can solve the gravitino overproduction problem by their distinctive shape of the potential.
  Furthermore, we found that this inflation model simultaneously
can  explain the observed baryon asymmetry through the non-thermal
 leptogenesis and is consistent with the WMAP results, that is, $n_s=0.951^{+0.015}_{-0.019}$
 and the negligible tensor to scalar ratio.
 }
\end{center}
\end{titlepage}

\setcounter{page}{2}

\section{Introduction}
Supersymmetric standard model has been considered
as the most promising candidate of the beyond standard model. In this
model, the large gauge hierarchy and the observed dark matter density
are explained and the gauge coupling unification is realized.
It is also considered that
the inflationary era \cite{Guth:1980zm} must exist in the very early universe.
Many cosmological problems such as the flatness and the horizon problems can
be solved by the presence of the inflationary era.
Furthermore, the observed almost scale invariant spectrum of CMB \cite{Spergel:2006hy}
is also explained by the flat potential of the inflaton.

However, it was recently found that many inflation models are not compatible
with the supersymmetric standard models,
since the gravitinos are overproduced by the inflaton decay \cite{Kawasaki:2006gs}.
Especially in the gravity mediation scenario, the situation is
disastrous, since the neutrality of the SUSY breaking field makes the
inflaton decay
rate into a pair of gravitinos larger.
Furthermore, the neutrality causes an overproduction of the SUSY breaking
fields which mainly decay into gravitinos \cite{Ibe:2006am}.
By the study of these effects, it was revealed that most of the known inflation models,
that is, chaotic, hybrid, topological and new inflation models
are disfavored in the gravity mediation scenario \cite{Kawasaki:2006gs,Ibe:2006am,Endo:2007sz,Endo:2006qk,Endo:2007ih,Ibe:2007si}. 

In this paper, we try to solve these problems without severe fine
tunings. As solutions, we have two alternatives in principal, that is, (i)
changing the hidden sector and (ii) constructing a successful inflation
model \footnote{Surely we have a third alternative in which an
exotic sector is added only for the solution. However, we do not consider
this alternative, since it does not seem to be a minimal solution.}.
In this paper, we focus on the second alternative, since the first alternative is not
effective for the generic gravity mediation scenarios and it was already
tried in \cite{Endo:2007cu}, which showed that the
gravitino problem can be avoided in a specific tuned gravity mediation
model.
We also examine whether the inflation model
is compatible with the observed red tilted spectrum
and accommodates the successful baryogenesis mechanism
which is a necessary ingredient in the present universe.
As a result, we find that inverted hybrid inflation models
\cite{Lyth:1996kt} can solve the all cosmological problems without any fine tunings.

The organization of this paper is as follows.
In section \ref{se:problem} we briefly explain the cosmological
problems of inflation models assuming that SUSY breaking is mediated by
Planck suppressed operators and gravitinos are unstable.
In section \ref{sse:polonyi}, we explain the gravitino overproduction
problem by the decay of the SUSY breaking field and a successful solution for
this problem.
We explain the gravitino production by the decay of inflaton
and other gravitino sources in section \ref{sse:inflatondecay} and
\ref{sse:other} respectively.
In section \ref{se:solution}, we study an inverted hybrid
inflation model and show that this inflation model solves the gravitino overproduction problems
and accommodates the leptogenesis scenario.
Finally, we summarize our results
 in section \ref{se:conclusion}.

\section{Cosmological Problems in Gravity Mediation} \label{se:problem}
It was recently revealed that the gravitinos can be overproduced by the
decay of the SUSY breaking field, inflaton and other sources in the gravity
mediation scenario.
Here we briefly explain the generality of this problem and summarize the constraints
which we study in section \ref{se:solution}.
See \cite{Ibe:2006am,Endo:2007sz} for details of this problem.

\subsection{Cosmological problem of SUSY breaking field}\label{sse:polonyi}

\subsubsection{Polonyi problem}

In the gravity mediation scenario, it is inevitable that
the SUSY breaking field has a very small mass $\sim O(100)$ GeV
and a very large amplitude $\sim O(M_\mathrm{pl})$ after the primordial
inflation, if the cut-off scale of the theory is the Planck scale.
These two features together cause a serious cosmological problem called
Polonyi problem \cite{Banks:1993en}.
Let us verify these two features and their implications for the cosmology.

In SUGRA, a ``Chirality-flipped mass matrix'' $M^2_{ij^*}$
at the vacuum with zero cosmological constant
can be written in terms of the total K\"ahler potential, $G=K+ \ln|W|^2$, as 
\begin{equation}
       M^2_{ij^*}
\equiv \left\langle
         \frac{\partial^2 V}{\partial \phi^i \partial \phi^{j^*}}
       \right\rangle
     = m_{3/2}^2
       \left\langle
         \nabla_i G_k \nabla_{j^*} G^k
       - R_{ij^*kl^*} G^k G^{l^*}
       + g_{ij^*}
       \right\rangle,
\end{equation}
where $m_{3/2}$ denotes the gravitino mass,
the subscripts $i$ denotes a derivative with respect to the field $\phi^i$,
and superscript is raised by $g^{ij^*}$
which is an inverse matrix of the K\"ahler metric $g_{ij^*}$
\footnote{
  In this paper, we take the unit with the reduced Planck scale
  $M_\mathrm{pl}\simeq 2.4\times 10^{18}$GeV equal to one,
  unless we explicitly denote.
}.
The curvature of the K\"ahler manifold $R_{ij^*kl^*}$ is defined by
$R_{ij^*kl^*} \equiv g_{ij^*kl^*} - g^{mn^*} g_{mj^*l^*}g_{n^*ik}$.
The covariant derivative is defined by
$\nabla_i G_j \equiv \partial_i G_j -\Gamma^k_{ij} G_k$,
where $\Gamma^i_{jk} \equiv g^{kl^*}g_{ijl^*}$ is the connection.

Potential minimization condition $\left<V_i \right>=0$ is also written as
\begin{equation}
  \left<G^j \nabla_i G_j +G_i \right> =0,
\end{equation}
which leads to an inequality:
\begin{equation}
  |G^jG_{ij}|=|g^{kl^*}g_{ijl^*}G^j G_k +G_i| < O(1).
  \label{eq:constraintGi}
\end{equation}
In the last inequality, we have used inequalities $G_i \lsim O(1)$
\footnote{
  The inequalities $G_i \lsim O(1)$ are required
  for the very small cosmological constant
  which is proportional to $(G^iG_i-3)$ in SUGRA.
  \label{Aug  8 14:20:06 2007}
}
and an assumption that a cut-off scale of the theory is the Planck scale, that is, 
all higher dimensional operators in the K\"ahler potential
are only divided by the Planck scale.

Using $G_Z \simeq \sqrt{3}$ and Eq.(\ref{eq:constraintGi}), inequalities
for the SUSY breaking field $Z$ are given by:
\begin{equation}
  |G_{iZ}| \lsim O(1),
\end{equation}
if cancellations do not occur in the left hand side of
Eq.(\ref{eq:constraintGi}).

Thus we can estimate an upper bound of the ``Chirality-flipped mass''
of the SUSY breaking field as
\begin{equation}
  M^2_{zz^*} \lsim m_{3/2}^2,
\end{equation}
where we have used $R_{ij^*kl^*}, g_{ij^*}\lsim 1$ which come from the
assumption that the cut-off scale is the Planck scale.
Moreover, ``Chirality-conserving mass'' $M^2_{zz}$
must be smaller than the ``Chirality-flipped mass'' $M^2_{zz^*}$,
since the potential becomes tachyonic otherwise.
Thus we have verified that the SUSY breaking field is always lighter than the
gravitino, when the cut-off scale of the theory is the Planck scale.

Now, let us consider the initial amplitude of the SUSY breaking field $Z$.
The amplitude is determined by the potential of $Z$ during the primordial inflation
which is controlled by the charge of the SUSY breaking field $Z$.
Here note that the SUSY breaking field must be neutral for any symmetry,
since the MSSM gaugino masses are given by operators
$\int d^2 \theta \frac{Z}{M_\mathrm{pl}} W^\alpha W_\alpha $,
where $W^\alpha$ denote the gauge field strength chiral superfields.

This neutrality of the SUSY breaking field implies that
we cannot forbid a linear term of the SUSY breaking field
in the K\"ahler potential:
\begin{equation}
  K(Z) = c_0^\dagger Z + c_0 Z^\dagger +|Z|^2 + \cdots , \label{eq:ztadpole}
\end{equation}
where $c_0$ is a dimensionful parameter
and expected to be of the order of the Planck scale.
Furthermore, even if some unknown mechanism at the Planck scale
suppresses the coefficient at the tree-level,
a linear term with $c_0 \gsim O(N_{g}/16 \pi^2)M_\mathrm{pl}$ is generated
through $1-$loop diagrams in which $N_g$ MSSM gauge multiplets circulate.
Thus we must consider that the coefficient of the linear term is
at least of order $0.1 M_\mathrm{pl}$, i.e. $|c_0| \gsim 0.1 M_\mathrm{pl}$.
Note that it contrasts with the FCNC constraints,
where the dangerous flavor changing operators are not induced,
if these operators are suppressed at the Planck scale.

By using this K\"ahler potential $K(Z)$
and  a potential of the inflaton $V_{\mathrm{inf}}$
which is nearly constant during the inflation,
the potential of the SUSY breaking field
during the inflation is approximately written as
\begin{eqnarray}
 V(Z)  & =& e^K |W|^2 \left( G_i G^i -3 \right) \cr
&\simeq& e^{K(Z)} V_\mathrm{inf} + m_Z^2 |Z|^2 \cr 
&\simeq& 3 e^{K(Z)} H_\mathrm{inf}^2 + m_Z^2 |Z|^2 \cr
&\simeq& 3 H_\mathrm{inf}^2(c_0^\dagger Z + c_0 Z^\dagger) +(3H_\mathrm{inf}^2+ m_Z^2)|Z|^2 + \cdots,
\end{eqnarray} 
where $m_Z$ denotes the mass of the SUSY breaking field and
$H_{\mathrm{inf}}$ is the Hubble constant during the primordial inflation.
In the third equality, we have used a Friedmann equation
during the inflation $V_\mathrm{inf} \simeq 3 H_\mathrm{inf}^2$.
Thus we can verify that the SUSY breaking field have a large initial
amplitude $\Delta Z \simeq c_0 \gsim 0.1 M_{\mathrm{pl}}$ after the
inflation and the universe is dominated by the SUSY breaking field
with the large initial amplitude.
Here we have assumed an inequality 
$H_\mathrm{inf} > m_{3/2} \simeq O(100)\mathrm{GeV} \gsim m_Z$, which is
satisfied in most inflation models.

In the gravity mediation scenario,
the SUSY breaking fields interact with the visible sectors
only by Planck suppressed operators.
Thus the decay time $\tau_Z \gsim M_\mathrm{pl}^2/m_Z^3 \gsim M_\mathrm{pl}^2/m_{3/2}^3$
is much longer than the timescale of the nucleosynthesis $\simeq O(1)$ sec
for the typical gravitino mass range $\simeq O(100)$GeV.
Such a late time decay of the SUSY breaking
field with too large number density spoils the prediction
of the Big Bang nucleosynthesis.
This serious cosmological problem in the gravity mediation scenario
is called Polonyi problem \cite{Banks:1993en}
\footnote{
  Note that the SUSY breaking field has serious cosmological problem,
  even if its mass is rather large
  $m_{Z}\simeq O(10)\mathrm{TeV}$ \cite{Nakamura:2007wr}.
}.

\subsubsection{Solution: Dynamical SUSY breaking and low scale inflation models}

The Polonyi problem seems to be solved,
if a condition $m_Z \gg m_{3/2}$ is satisfied
and the SUSY breaking field decays rapidly.
One way to realize this situation is
to violate the assumption that all higher dimensional operators are only
divided by the Planck scale.
For example, by introducing a higher dimensional operator
$|Z|^4/ M^{\prime 2} \gg |Z|^4 / M_\mathrm{pl}^2$ into the K\"ahler potential,
a large mass term for the SUSY breaking field is induced as
\begin{equation}
  V \supset \frac{|F_Z|^2}{1 - \frac{|Z|^2}{M^{\prime 2}} + \cdots}
    \sim    \frac{\Lambda^4}{M^{\prime 2}} |Z|^2 + \cdots
    \gg     m_{3/2}^2 |Z|^2,
\end{equation} 
where $\Lambda =\sqrt{F_Z} = 3^{1/4} \sqrt{m_{3/2}M_\mathrm{pl}}$
denotes the SUSY breaking scale.

We can have such a higher dimensional operator $|Z|^4 / \Lambda^2$
in the K\"ahler potential by nonperturbative effects,
when SUSY is dynamically broken
\footnote{
  There is Izawa-Yanagida-Intriligator-Thomas model
  \cite{Intriligator:1996pu}
  as an example of having such a higher dimensional operator.
}. Furthermore, the electroweak scale is explained by the dimensional
transmutation in this case.
Thus it seems that the Polonyi problem is solved
when the electroweak scale is realized by the dynamical SUSY breaking models.
However, in this solution we must constrain
the energy scale of the inflation model.

First, let us consider the case
$H_{\mathrm{inf}} \gsim \sqrt{m_{3/2}m_Z} \simeq 10^7 \mathrm{GeV}$
\footnote{
  This constraint depend on the SUSY braking model,
  although following discussions are not significantly changed.
  See \cite{Ibe:2006am} for a detailed discussion
  and a specific constraint in a model.
}.
For a significantly modified potential by the large Hubble constant, the initial amplitude of the SUSY breaking field
is larger than the dynamical scale $\Lambda$ right after the inflation.
As it is clear from the change of signature of the K\"ahler metric,
for such a large field value the effective K\"ahler potential
$K_{\mathrm{eff}} \simeq |Z|^2 - |Z|^4 / \Lambda^2 + \cdots$
is not valid and the potential is very flat around the initial field value.
Thus we suffer from the recurrence of the Polonyi problem for such a
large Hubble constant \cite{Ibe:2006am}.

Second, let us consider the case $H_{\mathrm{inf}} \lsim \sqrt{m_{3/2}m_Z}$.
For this case, the initial amplitude is given by
\begin{equation}
      \Delta Z 
    = \frac{3 H_\mathrm{inf}^2}{m_Z^2} c_0
\lsim \Lambda,
\end{equation}
which does not cause the problem discussed in the last paragraph.
However, there is a constraint on the yield of unstable gravitinos,
and this gives an upper bound on $H_\mathrm{inf}$. From the operator $|Z|^4 / \Lambda^2$, a dimensionless coupling
between the SUSY breaking field and the goldstinos arises as:
\begin{equation}
        \int d^4 \theta \frac{|Z|^4}{\Lambda^2}
\supset \frac{F_Z^\dagger}{\Lambda^2} Z^\dagger \psi_Z \psi_Z
      = Z^\dagger \psi_Z \psi_Z,
\end{equation}
which leads to a large decay width into gravitinos: 
$\Gamma(Z \rightarrow \psi_{3/2} \psi_{3/2}) \sim m_Z/8\pi$. 
For this decay width,
the SUSY breaking field mainly decays into two gravitinos,
whose lifetime is longer than the time scale of the nucleosynthesis.
The yield of the gravitinos produced by the decay of the SUSY breaking field is
\begin{equation}
  Y^\mathrm{Pol}_{3/2}
= \frac{T_R m_Z (\Delta Z)^2}{2 H_\mathrm{inf}^2}
= \frac{9 T_R H_\mathrm{inf}^2}{2m_Z^3} c_0^2,
  \label{eq:polonyicon}
\end{equation}
where $T_R$ denotes the reheating temperature.

The observed light elements abundances give the following constraint
on the yield of the unstable gravitinos
\cite{Kawasaki:2004yh, Kohri:2005wn}:
\begin{eqnarray}
          Y_{3/2}
& \lsim & Y_{3/2}^\mathrm{upper} \\
          Y_{3/2}^\mathrm{upper}
    & = & \left\{
            \begin{array}{rr}
              1 \times 10^{-16} - 5 \times 10^{-14}
              \textrm{\, for \,} m_{3/2} \simeq
            & 0.1-  1 \mathrm{TeV} \cr
              2 \times 10^{-14} - 5 \times 10^{-14}
              \textrm{\, for \,} m_{3/2} \simeq
            &   1-  3 \mathrm{TeV} \cr
              3 \times 10^{-14} - 2 \times 10^{-13}
              \textrm{\, for \,} m_{3/2} \simeq
            &   3- 10 \mathrm{TeV} \cr
            \end{array}
          \right.
          (B_h \simeq 10^{-3}),
          \label{eq:hadron2}
\end{eqnarray}
where $B_h$ denotes the hadronic branching ratio,
which is assumed to be $10^{-3}$ for conservative discussions in this paper.
These observational constraints lead to a constraint
on the Hubble scale during the inflation:
\begin{eqnarray}
           H_\mathrm{inf}
  \lsim    H_\mathrm{inf}^\mathrm{MAX}
& \equiv & 1.5 \times 10^6 \mathrm{GeV}
           \left( \frac{m_Z}{10^{11} \mathrm{GeV}} \right)^{3/2}
           \left( \frac{10^6 \mathrm{GeV}}{T_R} \right)^{1/2}
           \left( \frac{M_\mathrm{pl}}{c_0} \right)
           \left( \frac{Y_{3/2}^\mathrm{upper}}{10^{-14}} \right)^{1/2}
           \label{Aug  8 16:11:28 2007} \cr
& \lsim  & 1.5 \times 10^7 \mathrm{GeV}
           \left( \frac{m_{3/2}}{\mathrm{TeV}} \right)^{3/4}
           \left( \frac{10^6 \mathrm{GeV}}{T_R} \right)^{1/2}
           \left( \frac{M_\mathrm{pl}}{c_0} \right)
           \left( \frac{Y_{3/2}^\mathrm{upper}}{10^{-14}} \right)^{1/2}.
           \label{eq:hubbleconstraint}
\end{eqnarray}
In the last inequality, we have used an inequality
$m_Z \lsim \Lambda_\mathrm{dyn} \sim 4\pi \Lambda$
\footnote{
  Here, note that this constraint can not be loosened
  by decreasing the reheating temperature,
  since the yield of gravitinos produced through the inflaton decay
  becomes large for low reheating temperature.
  We will discuss this contribution in the next subsection.
},
\footnote{
  Here we should note that the upper bound
  is close to the assumed constraint
  $H_{\mathrm{inf}} \lsim \sqrt{m_{3/2} m_Z}$.
  We must care whether the initial amplitude is smaller
  than the effective cut-off scale $\Lambda$ in each SUSY breaking model,
  if the considering inflaton model has a energy scale
  close to the upper bound in Eq.(\ref{eq:hubbleconstraint}).
}.

This result shows that
SUSY chaotic inflation \cite{Murayama:1993xu, Kawasaki:2000yn},
SUSY topological inflation \cite{Izawa:1998rh, Kawasaki:2000tv, Kawasaki:2001as},
and SUSY (smooth) hybrid inflation models
\cite{Dvali:1994ms, Stewart:1994pt, Lazarides:1995vr, Linde:1997sj}
are disfavored,
since the Hubble scale of these inflation models
are determined from the observed anisotropy of CMB as:
\begin{eqnarray}
  H_\mathrm{inf}
= \left\{
    \begin{array}{rl}
      10^{14}    \mathrm{GeV} & \mbox{ for Chaotic Inflation}           \\
      10^{11-14} \mathrm{GeV} & \mbox{ for Topological Inflation}       \\
      10^{13-15} \mathrm{GeV} & \mbox{ for (smooth) Hybrid Inflation} 
    \end{array}
  \right.
\end{eqnarray}
On the contrary, some low scale inflation models
can satisfy the constraint Eq.(\ref{eq:hubbleconstraint}).
For this reason, New inflation model 
($H_\mathrm{inf} \gsim 10^{5.4}\mathrm{GeV}$) \cite{Izawa:1996dv}
and other low scale inflation models
seem to be favored than the high scale inflation models in the gravity
mediation scenario \cite{Ibe:2006am}.


\subsection{Gravitino overproduction through Inflaton direct decay} \label{sse:inflatondecay}

In this subsection, we review gravitino overproduction problem
caused by the inflaton perturbative direct decay
as another constraint in this paper
\footnote{
  In this topic, `\textit{inflaton}' denotes a scalar field
  whose coherent oscillation dominates the universe right after the inflation.
  Thus `\textit{inflaton}' in this topic is not necessarily
  the origin of the exponential expansion of the universe.
  For example, the waterfall fields
  in the Hybrid Inflation models correspond to this field.
}
\cite{Kawasaki:2006gs}.
In spontaneously broken Super Gravity, the VEV of the inflaton is slightly
shifted from that in the rigid case and mixing between inflaton and the SUSY
breaking fields is nonzero, even if they do not directly couple in
the K\"ahler or super potential. Thus the inflaton  F-term, that is,
the coupling with the gravitino is nonzero in SUGRA. Considering these
effects, the inflaton decay width into
the gravitinos is given by  \cite{Endo:2006zj}
\begin{eqnarray}
 \Gamma(\phi \rightarrow 2 \psi_{3/2}) &=&
\frac{|G_\phi^\mathrm{eff}|^2}{288 \pi} \frac{m_\phi^5}{m_{3/2}^2 M_\mathrm{pl}^2},
\end{eqnarray}
where $m_\phi$ is the inflaton mass.
And the effective coupling $G_{\phi}^\mathrm{eff}$ is given by \cite{Endo:2006tf}
\begin{eqnarray}
|G_\phi^\mathrm{eff}|^2 & \simeq & \left|\sqrt{3} g_{\phi z^*} \frac{m_Z^2}{\mathrm{Max}[m_\phi^2, m_Z^2]} \right|^2 \cr
&+& \left|\sqrt{3} (\nabla_\phi G_z) \frac{m_{3/2}m_z^2}{\mathrm{Max}[m_\phi^2,m_z^2]m_\phi}\right|^2 + \left|3 \frac{m_{3/2}m_\phi}{\mathrm{Max}[m_\phi^2,m_z^2]}\right|^2  \label{eq:gefforig}.
\end{eqnarray}

In the gravity mediation scenario,
the effective coupling is much larger than that in the other mediation scenarios.
In this scenario, the SUSY breaking field must be neutral for any symmetry,
as we explained in section \ref{sse:polonyi}.
For the neutrality of the SUSY breaking field,
the following mixing terms proportional to $c_i \sim O(1)$,
$i=1, \cdots$ can not be forbidden by any symmetry like a linear term in
Eq.(\ref{eq:ztadpole}). Thus the K\"ahler and super potential are written as
\begin{eqnarray}
 K & = & |\phi|^2 + |Z|^2 + |\phi|^2 (c_1^\dagger Z + c_1 Z^\dagger) + \cdots, \\
 W & = & W_\mathrm{hid} + W_\mathrm{inf} + \sum_{i=2} c_i W_\mathrm{inf}^{(i)} Z,
         \label{eq:coup in super}
\end{eqnarray} 
where $W_\mathrm{hid}$, $W_\mathrm{inf}$, $W_\mathrm{inf}^{(i)}$
denote hidden, inflaton superpotential and each operators in the
inflaton superpotential and the ellipses denote the other higher
dimensional operators which may be dismissed in the following discussions.

For this Lagrangian the effective coupling $G^\mathrm{eff}_\phi$ is
approximately written as
\begin{eqnarray}
           |G^\mathrm{eff}_\phi|^2
& \simeq & \left|
             \sqrt{3} g_{\phi z^\dagger}
             \frac{m_z^2}{\mathrm{Max}[m_\phi^2, m_z^2]}
           \right|^2 
         + \left|
             \sqrt{3} \left(\nabla_\phi G_z\right)
             \frac{m_{3/2}}{m_\phi}
             \frac{m_z^2}{\mathrm{Max}[m_\phi^2, m_z^2]}
           \right|^2 \cr
& \simeq & 3 |\phi|^2
           \left( \frac{m_z^2}{\mathrm{Max}[m_\phi^2, m_z^2]} \right)^2
           \sum_{i=1} O(1) c_i^2 \cr 
& \equiv & 3 |\phi|^2
           \left( \frac{m_z^2}{\mathrm{Max}[m_\phi^2, m_z^2]} \right)^2 C^2
           \label{eq:Geffective},
\end{eqnarray}
where $O(1)$ in the second line represent coefficients
depending on inflation models
\footnote{
  Here we neglected the third term in Eq.(\ref{eq:gefforig}),
  since the contribution is much smaller than the other contributions
  in Eq.(\ref{eq:Geffective}).
}.

The yield of the gravitinos produced through the inflaton decay
can be calculated by solving the Boltzmann equation.
The solution is approximately written as:
\begin{eqnarray}
           Y_{3/2}^\mathrm{inf} 
& \simeq & 2 \frac{\Gamma(\phi \rightarrow 2 \psi_{3/2})}{\Gamma_\phi}
             \frac{3 T_R}{4 m_\phi} \cr
& \simeq & 2.3 \times 10^{-6}
           \left( \frac{m_z^2}{\mathrm{Max}[m_\phi^2, m_z^2]} \right)^2 \cr
&        &
  \times   \left( \frac{\left<\phi \right>}{10^{15} \mathrm{GeV} } \right)^2
           \left( \frac{m_\phi}{10^{9} \mathrm{GeV} } \right)^4
           \left( \frac{10^7 \mathrm{GeV}}{T_R } \right)
           \left( \frac{\mathrm{TeV}}{m_{3/2} } \right)^2 C^2,
           \label{Aug  7 14:37:10 2007} 
\end{eqnarray} 
where $\Gamma_{\phi}$ denotes the main decay width of the inflaton field.

This quantity must be smaller
than the upper bound in Eq.(\ref{eq:hadron2}):
\begin{equation}
  Y_{3/2}^\mathrm{inf} \lsim Y_{3/2}^\mathrm{upper}, \label{eq:infcon}
\end{equation}
This constraint is so severe that
(smooth) Hybrid Inflation is again disfavored for this reason \cite{Endo:2007sz}.
Furthermore, New inflations model also become incredible
\footnote{
  Note that this constraint can not be avoided
  by decreasing the mass of SUSY breaking field,
  since the contribution of Eq.(\ref{eq:hubbleconstraint}) increases in that case.
}.
When Eq.(\ref{eq:hubbleconstraint}) and (\ref{eq:infcon})
are simultaneously considered in New inflation models,
severe fine tunings $c_i \lsim 10^{-4}$ $ (i= 0,1, \cdots)$ are required
for a constraint $Y_{3/2} < 10^{-14}$, even if the mass of the SUSY
breaking field $m_Z$ is appropriately tuned \cite{Ibe:2007si}.
Thus it seems that New inflation models are also disfavored
as well as chaotic, hybrid and topological inflation models.

\subsection{Other contributions} \label{sse:other}
Here, we mention contributions from the MSSM and hidden sector gauginos.

In SUGRA, all scalar fields including the inflaton field
couple with the hidden gauge super multiplets
by super K\"ahler-Weyl and $\sigma$-model anomalies \cite{Endo:2007ih}.
This coupling induces a large decay width of the inflaton
into the hidden gauge bosons and fermions,
when the following condition is not satisfied:
\begin{equation}
     m_\phi \lsim m^\mathrm{hid}_{1/2}
\simeq \Lambda_\mathrm{dyn} \simeq 4\pi \Lambda
\simeq 10^{12} \mathrm{GeV},
     \label{eq:gauginocon}
\end{equation}
where $m^\mathrm{hid}_{1/2} \simeq \Lambda_\mathrm{dyn}$
denotes the mass of the hidden gauge super multiplet.
Since these fields finally decay into gravitinos,
there can be a contribution comparable to Eq.(\ref{Aug  7 14:37:10 2007}),
if the constraint Eq.(\ref{eq:gauginocon}) is violated.
The gravitino yield is approximately given by \cite{Endo:2007sz}
\begin{align}
       Y_{3/2}^{\mathrm{hid}}
\simeq 9 \times 10^{-13} \xi
       \left( \frac{T_{R}}{10^{6} \mathrm{GeV}} \right)^{-1}
       \left( \frac{\langle \phi \rangle}{10^{15} \mathrm{GeV}} \right)^2
       \left( \frac{m_{\phi}}{10^{12} \mathrm{GeV}} \right)^2,
       \label{eq:hiddenyield}
\end{align}
where $\xi \simeq O(10^{-2})-O(10)$ is a constant depending on the hidden sector.
This yield must satisfy an inequality
\begin{align}
  Y_{3/2}^{\mathrm{hid}} < Y_{3/2}^{\mathrm{upper}}.\label{eq:hiddenyield2}
\end{align}
Thus the successful inflation model must satisfy the constraint
Eq.(\ref{eq:gauginocon}) or (\ref{eq:hiddenyield2}).

There is also another well known source of gravitinos,
that is, a contribution from thermal scatterings
of MSSM gluinos \cite{Weinberg:1982zq}.
This contribution $Y_{3/2}^\mathrm{th}$ is written as:
\begin{eqnarray}
           Y_{3/2}^\mathrm{th}
& \simeq & 1.9 \times 10^{-12}
           \left[
             1 + \left( \frac{m^2_{\tilde{g}_3}}{3 m_{3/2}^2} \right)
           \right]
           \left( \frac{T_R}{10^{10} \mathrm{GeV}} \right) \cr
&        &
  \times   \left[
             1 + 0.045 \ln \left( \frac{T_R}{10^{10} \mathrm{GeV} }\right)
           \right]
           \left[
             1 - 0.028 \ln \left( \frac{T_R}{10^{10} \mathrm{GeV} }\right)
           \right],
\end{eqnarray}
where $m_{\tilde{g}_3}$ is the gluino running mass evaluated at $\mu =T_R$.
This quantity also must satisfy a condition
\begin{equation}
  Y_{3/2}^\mathrm{th} \lsim Y_{3/2}^\mathrm{upper}, \label{eq:trcon}
\end{equation}
where $Y_{3/2}^\mathrm{upper}$ is the same as that in Eq. (\ref{eq:hadron2}).

Before closing this section,
we summarize the constraints for successful inflation model.
The gravitino sources and constraints are given as follows
\begin{description}
  \item[(i) The SUSY breaking field $Z$:]
    If the Hubble constant during the inflation is too large,
    the decay of the SUSY breaking field produces too many gravitinos.
    This gravitino source gives the constraint Eq.(\ref{Aug  8 16:11:28 2007})
    for inflation models.
  \item[(ii) Inflaton direct decay:]
    The gravitinos are also directly produced by the inflaton perturbative decay,
    when the inflaton field has non-vanishing VEV.
    Thus Eq.(\ref{eq:infcon}) must be satisfied in successful inflation models.
  \item[(iii) Gauge super multiplets in hidden sector:]
    If the inflaton field is heavier than the hidden gauge super multiplets,
    the decay of the inflaton produces these gauge bosons and fermions,
    whose decay can cause the gravitino overproduction.
    This overproduction can be avoided,
    if Eq.(\ref{eq:gauginocon}) or (\ref{eq:hiddenyield2}) is satisfied.
  \item[(iv) Thermal scatterings:]
    Thermal scatterings of gauginos in visible sector also produce the gravitinos.
    Since the gravitino yield is proportional to the reheating temperature,
    there is an upper bound for the reheating temperature
    which is given by Eq.(\ref{eq:trcon}).
\end{description}
In the gravity mediation scenario,
it has been revealed that most of the known inflation models are disfavored
on account of these four gravitino sources.
In the next section, we will try to find a inflation model
satisfying these constraints,
and check whether the baryogenesis can be accommodated.

\section{Solution to the Gravitino Overproduction problem}\label{se:solution}

\subsection{Inverted Hybrid inflation and its energy scales}
In this section, we study inflation models
consistent with the WMAP observation:
$\delta \rho / \rho = 1.9 \times 10^{-5}$, $n_s = 0.951^{+0.015}_{-0.019}$ \cite{Spergel:2006hy}
and the constraints argued in section \ref{se:problem}.

Inflation models consistent with the WMAP result
can be classified by their shapes of the potentials as follows
\cite{Alabidi:2006qa}:
\begin{equation}
V = \left\{
      \begin{array}{cll}
        \lambda \phi^n
      & n \ge 2
      & \mbox{Chaotic Inflation}, \\
        V_0 \left[ 1 - \left( \frac{\phi}{\mu} \right)^p \right]
      & \mathrm{p \le -0}
      & \mbox{Hybrid Inflation}, \\
        V_0 \left[ 1 - \left( \frac{\phi}{\mu} \right)^p \right]
      & \mathrm{p \ge 2}
      & \mbox{New, Topological  or Inverted Hybrid Inflation}.
      \end{array}
    \right.
\end{equation}
Although all these inflation models can explain the WMAP results,
these models except for ``inverted'' hybrid inflation models
are disfavored as we saw in section \ref{se:problem}.
Thus we focus on studying
whether there is an inverted hybrid inflation model
satisfying the constraints in section \ref{se:problem}
and generating the observed baryon asymmetry
\footnote{
  We are also interested in A-term inflation \cite{Lyth:2006ec},
  since this inflation model also suppress the gravitinos produced
  through the processes (i), (ii), (iii) and (iv).
  However, we do not focus on this possibility in this paper,
  since it seems for us that all the baryogenesis mechanisms
  except for electroweak baryogenesis cannot work and the initial field
  value of the inflaton must be fine tuned in this set up.
}.

\begin{figure}[tbp]
\begin{center}
\includegraphics{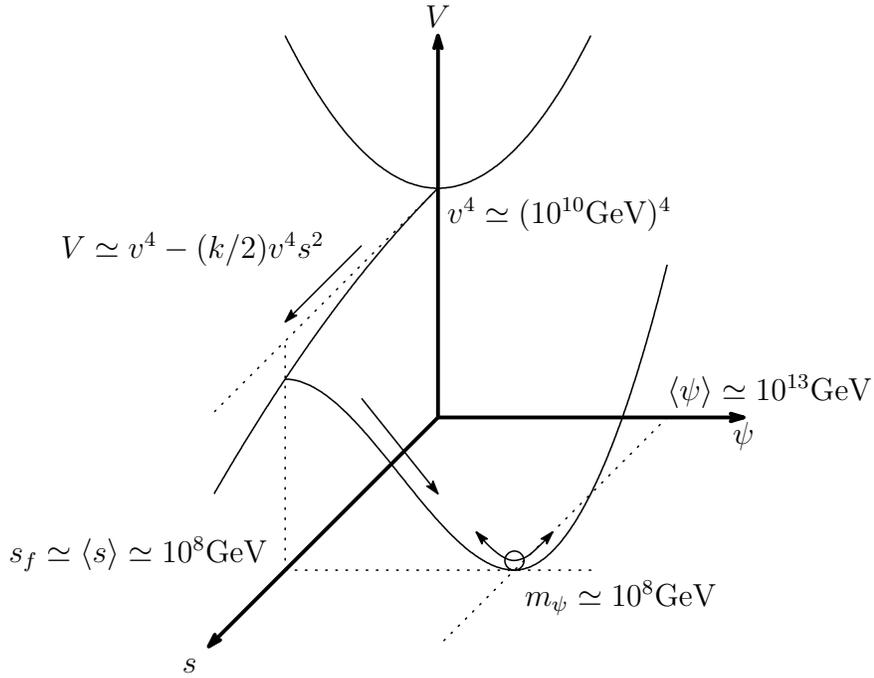}
\label{fig:potential}
\caption{
  This figure is an ideal potential of the inverted hybrid inflation model.
  In this potential the COBE normalization
  $V^{3/2} / V' \simeq v^{2} / (k s_{f}) \simeq 5 \times 10^{-4}$
  can be satisfied by
  $v \sim 10^{10}$ GeV, $s_{f} \simeq 10^8$ GeV and $k \simeq O(0.01)$
  ($k$ is also determined from the spectral index $n_s$).
  When the reheating temperature takes an appropriate value
  $T_R\simeq 10^6 \mathrm{GeV}$, we can also avoid the gravitino
 overproduction induced by the
 processes (i), (ii), (iii) and (iv).
}
\end{center}
\end{figure}

A successful Inverted Hybrid Inflation model
must have sufficiently low height of the potential,
a small VEV and a small mass of the waterfall field
to suppress the gravitinos produced through the processes (i), (ii) and (iii).
We expect that the constraints from (i), (ii), (iii), (iv),
and the WMAP observations
($n_s \simeq 1- 2 k = 0.951^{+0.015}_{-0.019}$ and
 $ V^{3/2}/ V'\simeq v^2 / k s_f \simeq 5.3 \times 10^{-4}$)
are satisfied by a potential like in Fig. \ref{fig:potential},
if the reheating temperature takes an appropriate value $\simeq 10^6$ GeV.
Here $k$ is related to a coefficient
of dimension 4 operator in the K\"ahler potential,
$v \equiv V^{1/4}$ represents the height of the potential
and $s_f$ denotes a field value of the inflaton at the end of the inflation.
We will explain the details of these parameters in a specific model.

This potential can be written as
\begin{eqnarray}
      V(s, \psi)
& = & (v^2 - \frac{\lambda}{4} \psi^2)^2
  +   (A \sqrt{\lambda} v - \frac{g}{\sqrt{2}} s)^2 \frac{\psi^2}{2} \cr
&   & {} - \frac{k}{2 M_\mathrm{pl}^2} v^4 s^2
  +   \frac{k^\prime}{2 M_\mathrm{pl}^2}v^4 \psi^2,
      \label{eq:invertedpotential}
\end{eqnarray}
where $s$ and $\psi$ are real scalar fields,
and the constants
$v\simeq O(10^{10}) \mathrm{GeV}$, $\lambda \simeq O(10^{-5})$,
$A > 1$, $g \simeq O(1)$, $k \simeq O(0.01)$ and $k^\prime$
are all real and positive
\footnote{
  It seems that we have fine tuned $\lambda$ in this model
  instead of $c_i$ parameters.
  However, notice that
  tunings of superpotential couplings are technically natural
  and the coupling $\lambda$ can be small
  by some symmetries in specific models.
  It is not the case for the $c_i$ parameters.
}.

In the inverted hybrid inflation model,
the inflaton $s$ and the waterfall field $\psi$
are set to the origin at first by some mechanism
\footnote{
  This mechanism may be finite temperature effects
  or Hubble mass induced by a foregoing (chaotic) inflation.
  Anyway such mechanisms do not change the following results.
}.
Then the inflaton field slowly rolls down to large field value,
decreasing the coefficient of the second term in
Eq.(\ref{eq:invertedpotential}) which stabilizes $\psi$ to the origin.
Finally the inflation era ends by the $\psi$'s water-falling,
when the inflaton $s$ reaches to the waterfall point $s = s_f$,
which is given by
\footnote{
  See \cite{Lyth:1998xn} for a review
  of slow roll inflation and its prediction.
}:
\begin{eqnarray}
  s_f = \frac{\sqrt{2 \lambda}}{g}(A - 1) v.
\end{eqnarray}

The value $s_{N_e}$ of the inflaton
corresponding to the $e$-fold number $N_e$ is given by
\begin{eqnarray}
       N_e
\simeq \int^{s_{N_e}}_{s_f} d s
       \frac{1}{M_\mathrm{pl}^{2}} \frac{V(s,0)}{V^\prime(s,0)}
\simeq \int^{s_{N_e}}_{s_f} d s
       \frac{v^4}{-k v^4 s}
     = \frac{1}{k} \ln \left( \frac{s_f}{s_{N_e}} \right),
\end{eqnarray}
where $V^\prime(s,0)$ denotes a derivative by a inflaton field $s$.
This leads to
\begin{eqnarray}
  s_{N_e}
= e^{-k N_e} s_f
= \frac{\sqrt{2 \lambda}}{g} (A - 1) v e^{-k N_e}.
\end{eqnarray}

Now let us determine the inflation scale $v$
as a function of the other parameters.
The amplitude of the primordial density fluctuations is given by
\begin{align}
       \frac{\delta \rho}{\rho}
\simeq \frac{1}{5 \sqrt{3} \pi M_\mathrm{pl}^3}
       \frac{V^{\frac{3}{2}} (s_{N_0})}{|V'(s_{N_0})|}
\simeq \frac{1}{5 \sqrt{3} \pi}
       \frac{v^6}{k v^4 s_{N_0} M_\mathrm{pl}},
\end{align}
where $s_{N_0}$ is the value of the inflaton field
at the epoch of the present horizon exit.
Thus the inflation energy scale $v = V^{\frac{1}{4}}$ is written as
\begin{equation}
  v
= k \,
  \frac{V^{\frac{3}{2}}}{V' M_\mathrm{pl}^2}
  \sqrt{2\lambda}
  \left( \frac{A - 1}{g} \right)
  e^{- k N_0}.
\end{equation}
Owing to the COBE normalization
\begin{align}
       \frac{1}{M_\mathrm{pl}^3}
       \frac{V^{\frac{3}{2}}(s_{N_0})}{|V'(s_{N_0})|}
\simeq 5.3 \times 10^{-4},
\end{align}
the scale $v$ is expressed as
\begin{align}
       v
\simeq 5.7 \times 10^{10} \mathrm{GeV}
       \times \mathcal{C}(k, N_0)
       \times \frac{A - 1}{g}
       \times \left( \frac{\lambda}{10^{-5}} \right)^{\frac{1}{2}},
       \label{eq:scale determination}
\end{align}
for e-fold $N_0 \simeq 40$ and $|k| \lsim O(0.01)$,
where $\mathcal{C}(k,N_0)$ is a function of order unity.

Then let us consider the spectral index $n_s$ and the tensor to scalar
ratio $r$. These values for the small field value
$s_{N_0} \ll M_\mathrm{pl}$ are approximately given by
\begin{eqnarray}
        n_s
 &\simeq&   1 + 2 M_\mathrm{pl}^2
       \left. \left(
         \frac{V^{\prime \prime}(s,0)}{V(s,0)}
       \right) \right|_{s = s_{N_0}}
\simeq 1 - 2k, \label{eq:spectral} \\
       r
& \simeq &  16 \times \frac{M_\mathrm{pl}^2}{2}
       \left. \left(
         \frac{V^\prime(s,0)}{V(s,0)}
       \right)^2 \right|_{s = s_{N_0}}
\simeq 8 k^2 \frac{s_{N_0}^2}{M_\mathrm{pl}^2} \ll 1
\end{eqnarray}
Thus the observed red tilted spectrum $n_s=0.951^{+0.015}_{-0.019}$ and
the negligible tensor to scalar ratio can be easily explained in the
inverted hybrid inflation models. It is in contrast with the other
hybrid inflation models.

The e-folding number $N_0$ corresponding
to the present horizon is also given by
\begin{align}
       N_0
\simeq 67
     + \frac{1}{3} \ln \frac{H_\mathrm{inf}}{M_\mathrm{pl}}
     + \frac{1}{3} \ln \frac{T_R}{M_\mathrm{pl}}
\simeq 67
     + \frac{1}{3} \ln \frac{v^2}{\sqrt{3} M_\mathrm{pl}^2}
     + \frac{1}{3} \ln \frac{T_R}{M_\mathrm{pl}},
       \label{eq:efold}
\end{align}
where $H_\mathrm{inf}$ denotes a Hubble scale at the horizon exit
and $T_R$ the reheating temperature.

By means of Eqs. (\ref{eq:scale determination}), (\ref{eq:spectral}) and
 (\ref{eq:efold}),
we can express inflation energy scale $v$ and the Hubble constant $H_\mathrm{inf}$
by the couplings $A$, $g$, $\lambda$ and the reheating temperature $T_R$.
For $T_R \simeq 10^6 \mathrm{GeV}$, the Hubble constant is approximately
given by
\begin{align}
       H_\mathrm{inf}
\simeq 1.7 \times 10^7 \mathrm{GeV}
       \times \lambda \left( \frac{A - 1}{g} \right)^2.
       \label{eq:hubble}
\end{align}
Thus, the constraint Eq.(\ref{eq:hubbleconstraint}) from the process (i)
is satisfied for small $\lambda$, which can be controlled by some symmetries.

\subsection{Reheating and Inverted hybrid inflation in SUGRA}

In this subsection, we study the other constraints
from the processes (ii), (iii) and (iv),
which are strongly related to the SUGRA effects and the reheating process.
Here we also take into account the baryogenesis mechanism
which is a very important process after the inflation.

For these purposes, we examine a following model
\footnote{
  Here we dismiss superpotential interaction terms
  like the third term in Eq.(\ref{eq:coup in super}),
  since the effects of these operators can be absorbed
  into the definition of $c_1$ in Eq.(\ref{eq:int in K}).
  See Eq.(\ref{eq:Geffective}) for this redefinition.
}:
\begin{eqnarray}
  K & = &  K^\mathrm{(eff)}_\mathrm{Pol} + K_\mathrm{inf} + K_\mathrm{int} \\
  W & = &  W^\mathrm{(eff)}_\mathrm{Pol} + W_\mathrm{inf}.
\end{eqnarray}
Each component in the K\"ahler potential is given by
\begin{eqnarray}
      K^\mathrm{(eff)}_\mathrm{Pol}
& = & c_0^\dagger Z
    + c_0 Z^\dagger
    + |Z|^2
    - \frac{1}{12}\frac{m_z^2}{m_{3/2}^2 M_\mathrm{pl}^2}|Z|^4
    + \cdots,
      \label{eq:poloK}\\
      K_\mathrm{inf}
& = & |\Psi|^2 + |S|^2 + |X|^2 + |Y|^2 +|N|^2\cr
&  &
    + (1+k) |S|^2 |X|^2
    +( 1 - k^\prime) |\Psi|^2 |X|^2 \cr
& &    - \frac{k_1}{4}|X|^4
    - k_2 |X|^2 |Y|^2
    - k_3 |X|^2 |N|^2
    + \cdots, \\
      K_\mathrm{int}
& = & |\Psi|^2 (c_1^\dagger Z + c_1 Z^\dagger)
    + \cdots
      \label{eq:int in K},
\end{eqnarray}
where $Z, \Psi, S, X, Y$ and $N$ are chiral superfields
and $k, k', k_1, k_2$ and $k_3$ are positive constants.
Here $N$ denotes the right-handed neutrino
and $Z$ the SUSY breaking field
which would have large mass $m_Z \gg m_{3/2}$
by the fourth operator in Eq.(\ref{eq:poloK})
produced by the strong dynamics.
The ellipses denote higher dimensional operators,
which may be neglected during the inflation
and the following reheating era.

The operators in the superpotential are given by
\begin{eqnarray}
      W_\mathrm{Pol}^\mathrm{eff}
& = & \sqrt{3} m_{3/2}M_\mathrm{pl} Z, \\
      W_\mathrm{inf}
& = & X \left(v^2 - \frac{\lambda}{2} \Psi^2 \right)
    + Y (A \sqrt{\lambda} v - g S) \Psi
    + \frac{h}{2} \Psi N^2,
\end{eqnarray}
where constants $v, \lambda, A, g$ and $h$
can be chosen to be real and positive by field redefinition
without loss of generality
\footnote{
  We note that the Lagrangian has $U(1)_R \times Z_2$ symmetry
  and the superpotential can include other operators allowed by this symmetry.
  However, we dismiss these operators in this paper,
  since our attention is
  not to produce a complete model in particle physics
  but to show the existence of a model
  satisfying the cosmological constraints.
}.

Then, the scalar potential in SUGRA is approximately given by
\begin{align}
  V
& \simeq \left|v^2 - \frac{\lambda}{2} \Psi^2 \right|^2
  + |A \sqrt{\lambda} v - g S|^2 |\Psi|^2 \cr
& - \frac{ k}{M_\mathrm{pl}^2} v^4 |S|^2
  + \frac{ k'}{M_\mathrm{pl}^2} v^4 |\Psi|^2 \cr
& + \left|
      \lambda X \Psi
    - Y(A \sqrt{\lambda} v - g S)
    - \frac{h}{2} N^2
    \right|^2
  + g^2 |Y|^2 |\Psi|^2
  + h^2 |\Psi|^2 |N|^2 \cr
& + \frac{k_1}{M_\mathrm{pl}^2} v^4 |X|^2
  + \frac{1 + k_2}{M_\mathrm{pl}^2} v^4 |Y|^2
  + \frac{1 + k_3}{M_\mathrm{pl}^2} v^4 |N|^2
  + \cdots,
\end{align}
where the terms in the second and fourth lines
are the Hubble mass terms induced by the SUGRA contributions.
The ellipses denote the other SUGRA contributions
which are not important in the following discussions.
In this potential, all the scalar fields except for 
$\psi \equiv \sqrt{2} \, \mathrm{Re} \, \Psi$ and 
$s \equiv \sqrt{2} \, \mathrm{Re} \, S$ remain in the origin
during the inflation and the following coherent oscillation era,
if all the scalar component of the chiral super multiplets
are set to the origin before the primordial inflation
\footnote{
  Such a situation can be realized by thermal effects
  or another inflation before the primordial inflation.
}.
Thus we can use this model as a realization
of the potential Eq.(\ref{eq:invertedpotential}).

The energy of the universe during the coherent oscillation era
is dominated by the waterfall field,
whose initial amplitude and the mass are given by
\begin{align}
    \langle \psi \rangle
& = \frac{2}{\sqrt{\lambda}} v 
    \simeq 1.7 \times 10^{13} \mathrm{GeV}
    \times \frac{A-1}{g}, \\
    m_\psi
& = \lambda \langle \psi \rangle
    \simeq 1.7 \times 10^8 \mathrm{GeV}
    \times \frac{A-1}{g}
    \left( \frac{\lambda}{10^{-5}} \right),
    \label{eq:psimass}
\end{align}
if $T_R \simeq 10^6 \mathrm{GeV}$ is realized.
After the oscillation,
the waterfall field mainly decays into two right-handed neutrinos,
if an inequality $m_\psi > 2 m_N = 2 h \langle \psi \rangle$ is satisfied. 

The decay width is approximately given by
\begin{align}
  \Gamma_{\psi} \simeq \frac{h^2}{16 \pi} m_\psi.
\end{align}
From this decay width the reheating temperature is approximately given by
\begin{align}
       T_R
\simeq \left(
         \frac{10}{g_* \pi^2} M_\mathrm{pl}^2 \Gamma_{\phi}^2
       \right)^\frac{1}{4}
\simeq 1.1 \times 10^6 \mathrm{GeV} \times
       \frac{3h}{\lambda}
       \left( \frac{A-1}{g} \right)^\frac{1}{2}
       \left( \frac{\lambda}{6 \times 10^{-6}} \right)^\frac{3}{2},
\label{eq:tr}
\end{align}
where $g_*= 228.75$ is the massless degrees of freedom in the MSSM. 

The baryon asymmetry is produced
by the decay of these right-handed neutrinos as:
\begin{align}
  \frac{n_B}{s}
& \simeq
  8.2 \times 10^{-11} 
  \left( \frac{T_R}{10^6 \mathrm{GeV}} \right) 
  \left( \frac{2 m_N}{m_\psi} \right) 
  \left( \frac{m_{\nu_3}}{0.05 \, \mathrm{eV}} \right)
  \frac{1}{\sin^2 \beta} \, \delta_\mathrm{eff} \cr
& \simeq
  8.2 \times 10^{-11} 
  \left( \frac{T_R}{10^6 \mathrm{GeV}} \right) 
  \left( \frac{2 h}{\lambda} \right) 
  \left( \frac{m_{\nu_3}}{0.05 \, \mathrm{eV}} \right)
  \frac{1}{\sin^2 \beta} \, \delta_\mathrm{eff},
\end{align}
which should be $n_B/s \simeq 8.7 \times 10^{-11}$
for the successful nucleosynthesis.
Here $m_{\nu_3}$ denotes the mass of the heaviest (active) neutrino,
which is generated by the see-saw mechanism \cite{seesaw}.
The phase $\delta_\mathrm{eff}$ is the effective CP phase
defined in \cite{Fukugita:1986hr}
and $\tan \beta$ is the ratio of the vacuum expectation values
of the up- and down-type Higgs bosons in the MSSM.
Thus we have found that the successful nucleosynthesis can be realized
when the following conditions are satisfied
\footnote{
  Note that the constraint Eq.(\ref{eq:trcon}) also must be satisfied.
}:
\begin{eqnarray}
  \lambda &\simeq& 6\times 10^{-6} \times \left( \frac{g}{A-1}\right)^\frac{1}{3},\\
 h &\simeq& 2 \times 10^{-6} \times \left( \frac{g}{A-1}\right)^{\frac{1}{3}},
\end{eqnarray}
where we have assumed $m_\psi / m_N =\lambda /h = 3 $ for the briefness.

\begin{figure}[tbp]
\begin{center}
\includegraphics{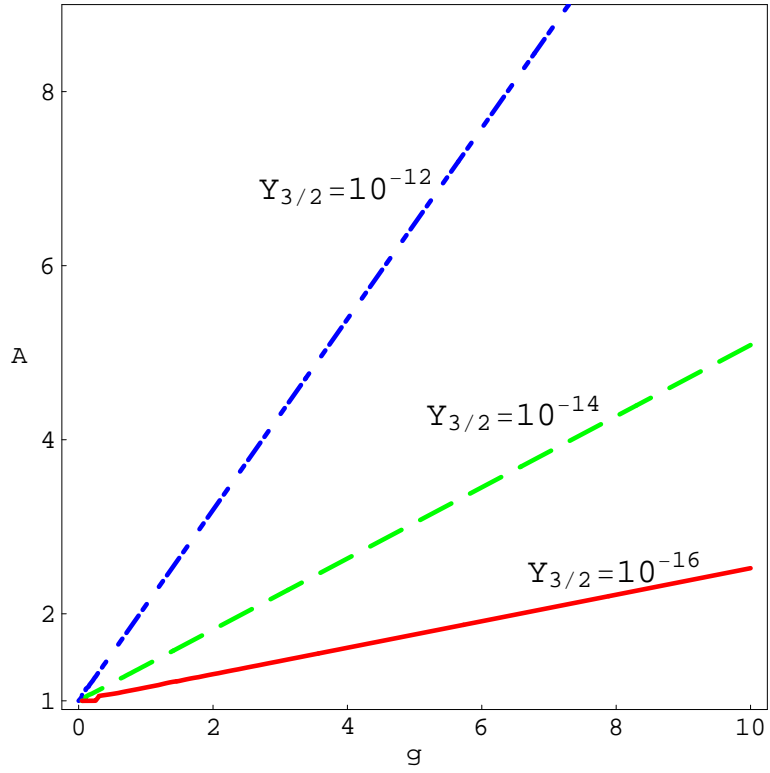}
\caption{
  The gravitino yield in the inverted hybrid inflation model is plotted.
  The (red) solid line denote $Y_{3/2}=10^{-16}$,
  (green) dashed line $Y_{3/2}=10^{-14}$
  and (blue) dash-dotted line $Y_{3/2}=10^{-12}$.
  In all the parameter region of this figure,
  the WMAP observation is reproduced
  and the sufficient baryon asymmetry is produced.
}
\label{CPY32}
\end{center}
\end{figure}

From the above discussions,
we can represent $H_\mathrm{inf}$, $m_{\psi}$ and $Y_{3/2}^\mathrm{inf}$
in terms of $g$, $A$ and $C$
by using the observed baryon asymmetry and the COBE normalization as:
\begin{align}
    H_\mathrm{inf}
  & \simeq
    \frac{\sqrt{V}}{\sqrt{3} M_\mathrm{pl}}
    \simeq
    \frac{v^{2}}{\sqrt{3} M_\mathrm{pl}}
    \simeq
    100 \, \mathrm{GeV}
    \left( \frac{A - 1}{g} \right)^{5/3}, \\
    m_{\psi}
& = \sqrt{\lambda} v
    \simeq 1.0 \times 10^{8} \mathrm{GeV}
    \left( \frac{A - 1}{g} \right)^{2/3}, \\
    Y_{3/2}^\mathrm{inf}
  & \simeq 2
    \, \frac{\Gamma_{3/2}}{\Gamma_{\psi}}
    \, \frac{3}{4}
    \, \frac{T_{R}}{m_{\psi}}
    \simeq
    6.5 \times 10^{-13}
    \left( \frac{A - 1}{g} \right)^{14/3} C^{2}.
\end{align}
In Fig.\ref{CPY32}, we have plotted the yield of the gravitinos for
$C=1$. From this figure and the equations above,
we see that this inverted Hybrid inflation model
can sufficiently suppress gravitinos produced though all processes,
simultaneously producing sufficient fluctuation of CMB
and the baryon asymmetry.
As a conclusion of this section, we have confirmed that the inverted
hybrid inflation model is consistent with the gravity mediation scenario.

\section{Summary} \label{se:conclusion}
The supersymmetric standard model is very attractive, since it can
explain the large hierarchy and the observed dark matter
density. However, it was discovered that the supersymmetric standard
model is not consistent with most of the inflation models, if the SUSY
breaking is mediated by Planck suppressed operators. Under these
circumstances, we have searched for successful inflation models in this paper.

Examining the gravitino production processes (i), (ii), (iii) and (iv)
in the gravity mediation scenario,
we have found that successful inflation models
must have sufficiently low hight of the potential,
an appropriate reheating temperature,
and a small mass and a small VEV of the inflaton at the true vacuum.
Studying these requirements and the WMAP results,
we have found that a particular inverted hybrid inflation model
is an example of such successful inflation models.
Furthermore, we have found that this inflation model
accommodates the see-saw mechanism
and produces a sufficient baryon asymmetry by the leptogenesis mechanism.
We consider that appropriate inflation models
including this inverted hybrid inflation model
will become interesting,
when the gravity mediation scenario is confirmed
in the future accelerator experiments.

\section*{Acknowledgments}
Y.S. thanks the Japan Society for the Promotion of Science for
financial support.
We thank T.T. Yanagida for useful suggestions.

\end{document}